\documentclass[10pt,aps,prd,twocolumn,floats,floatfix,showpacs,superscriptaddress,nofootinbib]{revtex4-2}
\usepackage[utf8]{inputenc} 

\usepackage{graphicx,mathtools,amssymb,amsmath,amsthm,amsfonts,epsfig,epsf,bm}
\usepackage[outdir=./]{epstopdf}
\usepackage[linktocpage]{hyperref}
\hypersetup{hidelinks}
\hypersetup{pdfstartview=}
\hypersetup{
    colorlinks=true,
    linkcolor=blue,
    filecolor=magenta,      
    urlcolor=BrickRed,
    citecolor=BrickRed,
}
\usepackage[usenames]{color}
\usepackage{tensor}
\usepackage{csquotes}
\usepackage{mathrsfs}
\usepackage{autobreak}
\usepackage{soul}

\usepackage{slashed}

\def\be{\begin{equation}}
\def\ee{\end{equation}}
\def\beq{\begin{eqnarray}}
\def\eeq{\end{eqnarray}}

\def\cR{{\cal R}}
\def\fR{f({\cal R})}

\newcommand*\DAlembert{\mathop{}\!\mathbin\Box}

\usepackage[dvipsnames]{xcolor}

\begin{document}

\title{Incompatibility of gravity theories with auxiliary fields with the standard model}

\author{\textbf{Giulia Ventagli}}
\affiliation{Nottingham Centre of Gravity, University of Nottingham,
University Park, Nottingham NG7 2RD, United Kingdom}
\affiliation{School of Mathematical Sciences, University of Nottingham,
University Park, Nottingham NG7 2RD, United Kingdom}
\affiliation{CEICO, Institute of Physics of the Czech Academy of Sciences, Na Slovance 2, 182 21 Praha 8, Czechia}

\author{\textbf{Paolo Pani}}
\affiliation{Dipartimento di Fisica, ``Sapienza'' Universit\`a di Roma, Piazzale 
Aldo Moro 5, 00185, Roma, Italy}

\author{\textbf{Thomas P.~Sotiriou}}
\affiliation{Nottingham Centre of Gravity, University of Nottingham,
University Park, Nottingham NG7 2RD, United Kingdom}
\affiliation{School of Mathematical Sciences, University of Nottingham,
University Park, Nottingham NG7 2RD, United Kingdom}
\affiliation{School of Physics and Astronomy, University of Nottingham,
University Park, Nottingham NG7 2RD, United Kingdom}


\begin{abstract}
Theories of gravity with auxiliary fields are of particular interest since they are able to circumvent Lovelock's theorem while avoiding the introduction of new degrees of freedom. This type of theories introduces derivatives of the stress-energy tensor in the modified Einstein equation. This peculiar structure of the field equations was shown to lead to spacetime singularities on the surface of stars. Here we focus on yet another problem afflicting gravity theories with auxiliary fields. We show that such theories can generically introduce parametrically large deviations to the Standard Model unless one severely constrains the parameters of the theory, preventing them to produce significant phenomenology at
large scales. We first consider the specific case of Palatini $f({\cal R})$ gravity, to clarify the results previously obtained in Ref.~\citep{Flanagan:2003rb}. We show that the matter fields satisfy the Standard Model field equations which reduce to those predicted by general relativity in the local frame only at tree level, whereas at higher orders in perturbation theory they are affected by corrections that percolate from the gravity sector regardless of the specific $f({\cal R})$ model considered. Finally, we show that this is a more general issue affecting theories with auxiliary fields connected to the same terms responsible for the
appearance of surface singularities.
\end{abstract}

\maketitle

\section{Introduction}

Although general relativity (GR) is widely accepted as a classical theory of gravity, there is both theoretical and experimental motivation to go beyond GR.  On the theoretical side,  GR is not renormalizable and does not offer a quantum description of gravitation. Furthermore, it predicts the appearance of spacetime singularities, in the vicinity of which GR itself breaks down.
On the experimental side, GR is challenged by the rich and overwhelming astronomical evidence for the existence of a large amount of unseen dark matter~\cite{Zwicky:1933gu,Bertone:2004pz,Turner:1999kz} and dark energy~\cite{SDSS:2005xqv, SupernovaSearchTeam:2004lze,WMAP:2006bqn,Carroll:2000fy}. 

To go beyond GR, one  needs to evade Lovelock's uniqueness theorem~\cite{Lovelock:1971yv,Lovelock:1972vz} by abandoning one of its assumptions, namely introduce extra dimensions, allow for higher-order equations, break diffeomorphism invariance, explicitly include additional fields, or abandon the idea that gravity can be described by a field theory whose equations involve local differential operators~\cite{Berti:2015itd}. Although dropping any one of these assumptions could lead to a very different quantum gravity theory, if one focuses on a local, classical effective field theory (as a low-energy limit) dropping any of the assumptions leads to the same outcome~\cite{Sotiriou:2014yhm}: additional fields. Typically, these fields will be dynamical and would propagate. Searches for such new fields are one of the main ways to test GR and have so far not given any detection. The dynamical behavior of these additional fields has also given strong theoretical bounds on deviation from GR, as they are prone to the Ostrogradski instability~\cite{Woodard:2006nt}.

It is then particularly interesting to consider theories with \emph{nondynamical} extra fields. Such theories circumvent Lovelock's theorem without adding any extra propagating degrees of freedom. The extra fields are instead auxiliary, {\em i.e.}~one does vary the action with respect to them but they can be determined algebraically from the field equations.\footnote{Note that auxiliary fields differ from background fields, as the latter are kept fixed when varying the  action.} In Ref.~\cite{Pani:2013qfa}, it was shown that, once auxiliary fields have been eliminated, one obtains the following field equations
\begin{equation}\label{eq:gravAux}
    G_{\mu\nu}+\Lambda g_{\mu\nu}=\kappa^2 T_{\mu\nu}+S_{\mu\nu},
\end{equation}
where $\kappa^2=8\pi G$ (henceforth we use $\hbar=c=1$ units), $T_{\mu\nu}$ is the matter stress-energy tensor and $S_{\mu\nu}$ is a divergence-free tensor that depends on the matter fields, the metric, and their derivatives. The precise form of $S_{\mu\nu}$ will depend on the specific form of the auxiliary field and how it enters the Lagrangian. However, a characteristic feature of this kind of theories is that $S_{\mu\nu}$ contains derivatives of the stress-energy tensor, and hence the theory inevitably includes higher-order derivatives of the matter fields in the field equations. As a consequence, the metric is overly sensitive to abrupt discontinuities in the matter energy density leading to spacetime singularities, e.g. on the surface of stars. This was explicitly shown for specific classes of gravity with auxiliary fields, first for Palatini $f({\cal R})$ gravity in Refs.~\citep{Barausse:2007pn,Barausse:2007ys} and more recently for Eddington-inspired Born-Infeld gravity~\cite{Pani:2012qd}.

Palatini $f({\cal R})$ gravity is perhaps the most characteristic and well-studied example of gravity with an auxiliary field~\cite{Vollick:2003aw,Flanagan:2003iw,Capozziello:2004vh,Olmo:2004hj,Sotiriou:2005hu,Sotiriou:2005xe,Sotiriou:2005cd,Carroll:2006jn,Sotiriou:2006hs,Exirifard:2007da,Li:2007xw,Allemandi:2004ca,Sotiriou:2005hu,Capozziello:2007ec,Sotiriou:2008rp,DeFelice:2010aj,Vitagliano:2010pq}. It became popular as a potential explanation of the accelerated cosmic expansion~\cite{SDSS:2005xqv, SupernovaSearchTeam:2004lze,WMAP:2006bqn}. It consists of a generalization of the Einstein-Hilbert action by allowing the Lagrangian to be a generic function of the Ricci scalar ${\cal R}$, while also assuming that the metric and the connection are independent variables. By doing so, the Ricci tensor and, consequently, the Ricci scalar present in the action are constructed with the independent connection. A crucial assumption in this formalism is that the matter action does not contain the independent connection~\cite{Sotiriou:2006sr}.

The issue with surface singularities is not the only problem with Palatini $f({\cal R})$ gravity.\footnote{Note that it was also recently shown in Refs.~\cite{Dioguardi:2021fmr,Dioguardi:2022oqu,Dioguardi:2023jwa} how the shape of the $f({\cal R})$ function and the matter potential are strongly related to each other; as a consequence not all possible configurations allow for a real solution of the auxiliary field equation, and higher-order Palatini $f({\cal R})$ produces divergences in the kinetic terms of matter.} While the theory  satisfies the weak equivalence principle, in the sense that the matter couples minimally to the metric and is not coupled to the independent connection, eliminating this connection from the field equations introduces matter corrections to them in perturbation theory. This was pointed out in  Ref.~\citep{Flanagan:2003rb}, where a dynamical equivalent scalar-tensor theory was used to show  that a specific model of Palatini $f({\cal R})$ gravity exhibits such corrections and that this is in conflict with particle-physics observations. There has been debate about this result in the literature~\citep{Vollick:2003ic,Flanagan:2004bz,Iglesias:2007nv} and, in particular, about whether it is specific to the choice of 
conformal frame~\citep{Vollick:2003ic}.

In this paper, we revisit this issue and attempt to clarify it and demonstrate that the conflict with the Standard Model is indeed there and it is not specific to the model studied in Ref.~\citep{Flanagan:2003rb}, or  even Palatini $f({\cal R})$ gravity. It is instead more generic to theories with auxiliary fields and indeed is linked to the same terms that are responsible for the appearance of surface singularities in such theories. For what regards the confusion between different conformal frames, it appears to be due to the fact that the effect appears at different orders in perturbation theory in different conformal frames, as we discuss in more detail below.

The paper is organized as follows. In Sec.~\ref{Sec:Palatini}, we give a general review of Palatini $f({\cal R})$ gravity. Then, in Sec.~\ref{Sec:ScalTens}, we recast Palatini $f({\cal R})$ gravity as a scalar-tensor theory, and we report the results obtained in Ref.~\citep{Flanagan:2003rb}. Section~\ref{Sec:MatterFields} is devoted to the perturbative analysis of the field equations, and we present our results for the case of a massless scalar field for a specific model of Palatini $f({\cal R})$ gravity. In Sec.~\ref{Sec:AuxiliaryFields}, we generalize these results to the broader class of theories presented in Ref.~\citep{Pani:2013qfa}. Finally, we conclude in Sec.~\ref{Sec:Discussion}.


\section{Palatini $f({\cal R})$ gravity}\label{Sec:Palatini}

The action for Palatini $f({\cal R})$ gravity takes the form 
\begin{equation}
\label{action}
S=\frac{1}{2 \kappa^2}\int d^4x \sqrt{-g} \left[f(\mathcal{R}) + {\cal L}_M(g_{\mu\nu},\psi)\right],
\end{equation}
where $g$ is the determinant of the metric $g_{\mu\nu}$ and $\mathcal{R}=g^{\mu\nu}\mathcal{R}_{\mu\nu}$, where $\mathcal{R}_{\mu\nu}$ is the Ricci tensor constructed with the independent connection $\Gamma^\rho_{\mu\nu}$. As we mentioned before, in the Palatini approach, the matter Lagrangian ${\cal L}_M$ generically depends on the matter fields $\psi$ and on the metric tensor, but not on the independent connection. This implies that the parallel transport is defined only by the Levi-Civita connection~\citep{Sotiriou:2008rp}, and that the stress-energy tensor, $T_{\mu\nu}=\frac{-2}{\sqrt{-g}}\frac{\delta \mathcal{L}_M}{\delta g^{\mu\nu}}$,  is divergence-free with respect to the metric covariant derivative. Thus, the above theory is expected to satisfy the weak equivalence principle.

The field equations stemming from action~\eqref{action} read~\citep{Sotiriou:2008rp}
\begin{eqnarray}
f'(\mathcal{R})\mathcal{R}_{\mu\nu}-\frac{1}{2} g_{\mu\nu}f(\mathcal{R})&=&\kappa ^2 T_{\mu\nu}, \label{fieldeq1}\\
\bar{\nabla}_{\rho} (\sqrt{-g} f'(\mathcal{R}) g^{\mu\nu}) &=&0\,, \label{fieldeq2}
\end{eqnarray}
where we have defined the covariant derivative constructed with the independent connection as $\bar{\nabla}_\rho$.
In the special case $f(\mathcal{R})=\mathcal{R}$, Eq.~\eqref{fieldeq1} reduces to Einstein's equations, whereas Eq.~\eqref{fieldeq2} becomes the definition of the Levi-Civita connection for $\Gamma^\rho_{\mu\nu}$.

Taking the trace of Eq.~\eqref{fieldeq1} yields an \emph{algebraic} equation between ${\cal R}$ and $T\equiv T_{\mu\nu}g^{\mu\nu}$,
\begin{equation}
f'(\mathcal{R})\mathcal{R}-2f(\mathcal{R})=\kappa^2 T. \label{algebraic}
\end{equation}
Thus, the Ricci scalar is determined in terms of the matter fields, i.e. $\mathcal{R}=\mathcal{R}(T)$. In particular, when $T=0$ the above equation reduces to $f'(\mathcal{R})\mathcal{R}-2f(\mathcal{R})=0$, and $\mathcal{R}$ has to be a constant defined by the root of this equation. Replacing this in Eq.~\eqref{fieldeq1}, one effectively solves GR with an effective cosmological constant. 

Finally, it is useful to define a metric conformal to $g_{\mu\nu}$,
\begin{equation}
\label{conformal}
h_{\mu\nu}\equiv f'(\mathcal{R})g_{\mu\nu}.
\end{equation} 
Using this relation, Eq.~\eqref{fieldeq2} becomes the definition of the Levi-Civita connection of $h_{\mu\nu}$, $\Gamma^\rho_{\mu\nu}=\frac{1}{2}h^{\rho\sigma}(\partial_\mu h_{\nu\sigma}+\partial_\nu h_{\mu\sigma}-\partial_\sigma h_{\mu\nu})$. Substituting Eq.~\eqref{conformal}, the independent connection can be written in terms of $\mathcal{R}$ and of the metric $g_{\mu\nu}$. Therefore, since Eq.~\eqref{algebraic} relates $\mathcal{R}$ to $T$, one can completely eliminate $\Gamma^{\rho}_{\mu\nu}$ from the field equations and rewrite the latter only in terms of the metric and of the matter fields. Indeed, first note that the Ricci tensor transforms under the conformal transformation~\eqref{conformal} as
\begin{equation}
\label{ricciConf}
\begin{split}
\mathcal{R}_{\mu\nu}=R_{\mu\nu}&+\frac{3}{2}\frac{1}{(f'(\mathcal{R}))^2}\nabla_\mu f'(\mathcal{R}) \nabla_\nu f'(\mathcal{R})\\&-\frac{1}{f'(\mathcal{R})}\left( \nabla_\mu \nabla_\nu +\frac{1}{2}g_{\mu\nu}\Box \right)f'(\mathcal{R})\,,
\end{split}
\end{equation}
where $R_{\mu\nu}={R^\alpha}_{\mu\alpha\nu}$ and ${R^\alpha}_{\mu\beta\nu}$ is the Riemann tensor constructed with the Levi-Civita connection.
Since the Ricci scalars constructed with the independent connection and the Levi-Civita connection are, respectively, defined as $\mathcal{R}=g^{\mu\nu}\mathcal{R}_{\mu\nu}$ and ${R}=g^{\mu\nu}{R}_{\mu\nu}$, we contract Eq.~\eqref{ricciConf} with the metric finding
\begin{equation}
\label{ricciScalConf}
\begin{split}
\mathcal{R}=R&+\frac{3}{2}\frac{1}{(f'(\mathcal{R}))^2}\nabla_\mu f'(\mathcal{R}) \nabla^\mu f'(\mathcal{R})\\&-\frac{3}{f'(\mathcal{R})}\Box f'(\mathcal{R}).
\end{split}
\end{equation}
Thus, we can rewrite the field equation~\eqref{fieldeq1} as
\begin{equation}
\label{eqfieldMod}
\begin{split}
G_{\mu\nu} & =\frac{\kappa^2}{f'}T_{\mu\nu}-\frac{1}{2}\frac{1}{f'}g_{\mu\nu}\left( \mathcal{R}-\frac{f}{f'}  \right)\\
    & +\frac{1}{f'}(\nabla_\mu\nabla_\nu f'-g_{\mu\nu}\square f')\\
    &-\frac{3}{2}\frac{1}{f'^2}\bigg[ (\nabla_\mu f')(\nabla_\nu f')\\
    &-\frac{1}{2}g_{\mu\nu}(\nabla_\lambda f')(\nabla^\lambda f') \bigg],
\end{split}
\end{equation}
where once again we stress that this is a function of only the metric and the matter fields, since, due to Eq.~\eqref{algebraic}, we have $\mathcal{R}=\mathcal{R}(T)$ and $f(\mathcal{R})=f(\mathcal{R}(T))$.
Note that the fact that we can eliminate $\Gamma^{\lambda}_{\mu\nu}$ in terms of the metric and the matter fields indicates that, in Palatini $f({\cal R})$ gravity, the independent connection is an \emph{auxiliary} field and does not propagate any additional degree of freedom.
In the specific case when $f(\mathcal{R})=\mathcal{R}$, we find from Eq.~\eqref{conformal} that $h_{\mu\nu}=g_{\mu\nu}$, since $f'=1$. Therefore, $\mathcal{R}_{\mu\nu}=R_{\mu\nu}$, $\mathcal{R}=R$, and Eq.~\eqref{eqfieldMod} reduces to the Einstein field equation, showing once again that the Palatini formalism applied to the Einstein-Hilbert action yields GR.

The differential structure of Eq.~\eqref{eqfieldMod} is clearly a cause for concern. 
This field equation is a second-order partial differential equation in the metric, as in GR, but at the same time it includes up to second derivatives of $f'(\mathcal{R})$, and consequently of \emph{T}, since from Eq.~\eqref{algebraic} we have $\mathcal{R}=\mathcal{R}(T)$. Note that the matter action usually contains derivatives of the matter field $\psi$, so that one has $T=T(\psi,\partial_\mu \psi)$; hence Eq.~\eqref{eqfieldMod} contains up to third-order derivatives of the matter field. In GR, the higher differential order in the metric with respect to the differential order in the matter fields guarantees that the metric comes as an integral over the matter fields, so that any discontinuities in the matter are ``smoothed out" and not inherited by the metric. However, in Palatini $f({\cal R})$ gravity this does not happen. Indeed, since the differential order in the matter fields is higher than in the metric, the latter is no longer an integral over the matter fields, but it is related to them and their derivatives. For example, the problem of surface singularities discussed in Ref.~\citep{Barausse:2007pn,Barausse:2007ys} is related to this mechanism: a discontinuity in the matter fields or in their derivatives can lead to curvature singularities.

We stress that while Palatini $f({\cal R})$ gravity exemplifies this peculiar differential structure, the latter is a more general feature of gravity theories with auxiliary fields. As we have mentioned, this is due to the fact that the ``sourcelike'' term $S_{\mu\nu}$ in Eq.~\eqref{eq:gravAux} contains a derivative of the stress-energy tensor.

In this paper, we focus on potential deviations from the Standard Model of Palatini $f({\cal R})$ gravity, and more in general of gravity theories with auxiliary fields. We will discuss this issue after we introduce in the next section the scalar-tensor formulation of action~\eqref{action}, where the auxiliary field is simply represented by a scalar field.


\section{Palatini $f({\cal R})$ gravity and scalar-tensor formulation}\label{Sec:ScalTens}

%
Palatini $f({\cal R})$ gravity can be cast in the form of a scalar-tensor theory in the presence of a nondynamical scalar field, obtaining~\citep{Flanagan:2003iw,Carroll:2006jn,Capozziello:2007ec,Sotiriou:2008rp,DeFelice:2010aj}
\begin{equation}
\label{actionST}
\begin{split}
S=&\frac{1}{2 \kappa^2}\int d^4x \sqrt{-g}\,\left( \phi R+\frac{3}{2\phi}\partial_\mu \phi \partial^\mu \phi-V(\phi) \right) \\ &+ S_M(g_{\mu\nu},\psi),
\end{split}
\end{equation}
where we have defined the scalar field $\phi=f'(\mathcal{R})$ and the scalar potential $V(\phi)=\mathcal{R}(\phi)\phi-f(\mathcal{R}(\phi))$. Note that in the above equation $R$ is the Ricci scalar constructed from the metric $g_{\mu\nu}$, as in the standard definition of scalar-tensor theories. Action~\eqref{actionST} is equivalent to a Brans-Dicke theory in the Jordan frame with $\omega=-3/2$ and a potential $V$. 

Varying the action with respect to the metric and the scalar field yields
\begin{equation}
\label{eq:Jordan}
\begin{split}
G_{\mu\nu} & =\frac{\kappa^2}{\phi}T_{\mu\nu}-\frac{3}{2\phi^2}\left( \nabla_\mu\phi\nabla_\nu\phi-\frac{1}{2}g_{\mu\nu}\nabla_\lambda\phi\nabla^\lambda\phi \right)\\
     & +\frac{1}{\phi}(\nabla_\mu\nabla_\nu\phi-g_{\mu\nu}\square\phi)-\frac{V}{2\phi}g_{\mu\nu},
\end{split}
\end{equation}
\begin{equation}
\label{72}
\square\phi=\frac{\phi}{3}(R-V')+\frac{1}{2\phi}\nabla_\lambda\phi\nabla^\lambda\phi.
\end{equation}
Taking the trace of Eq.~\eqref{eq:Jordan}, we can eliminate the Ricci scalar in Eq.~\eqref{72}, obtaining an algebraic relation,
\begin{equation}
\label{eq:potential}
2V(\phi)-\phi V'(\phi)=\kappa^2 T\,,
\end{equation}
that determines the (auxiliary) field $\phi$ in terms of $T$. 
Note that Eq.~\eqref{eq:Jordan} and Eq.~\eqref{eq:potential} are equivalent respectively to Eq.~\eqref{eqfieldMod} and to the algebraic relation Eq.~\eqref{algebraic}.

Finally, action~\eqref{actionST} can be written in the Einstein frame as
\begin{equation}
\label{actionE}
\tilde{S}= \int d^4x \sqrt{-\tilde{g}}\left[ \frac{\tilde{R}}{2 \kappa^2}-U(\phi) \right]+S_M(\phi^{-1}\tilde{g}_{\mu\nu},\psi_m),
\end{equation}
where $U(\phi)=\frac{V(\phi)}{2\kappa^2 \phi^2}$ and $\tilde{g}_{\mu\nu}=f'(R)g_{\mu\nu}\equiv \phi g_{\mu\nu}$.  (Henceforth, tilded quantities will refer to the Einstein frame.) In this frame, the field equations read
\begin{eqnarray}
\tilde{G}_{\mu\nu}&=&\kappa^2 \tilde{T}_{\mu\nu}-\kappa^2\tilde{g}_{\mu\nu}U(\phi)\,, \label{eqEF1} \\
U'(\phi)&=&-\frac{1}{2}\tilde{T}\phi^{-1}\,, \label{eqEF2}
\end{eqnarray}
and the scalar field is manifestly auxiliary.

Let us now briefly review the results of Ref.~\citep{Flanagan:2003rb}, where the theory is studied in the Einstein frame choosing the specific model corresponding to $f(\mathcal{R})=\mathcal{R}-\mu^4/\mathcal{R}$, where one takes $\mu$ to be a mass scale of order the Hubble scale, i.e. $\mu \propto H_0$ , in order to describe the present acceleration of the universe. They considered the case where the matter action is described by the Dirac action.
The auxiliary scalar field $\phi$ is canonically renormalized as
\begin{equation}
\label{Phi}
\Phi=\frac{\sqrt{6}}{2\kappa}\log \phi,
\end{equation}
so that the total action written in the Einstein frame is
\begin{equation}
\label{actionFlan}
\begin{split}
\tilde{S}=\int d^4x \sqrt{-\tilde{g}} \biggl( \frac{\tilde{R}}{2\kappa^2}-U(\Phi)+&ie^{2\alpha(\Phi)}\bar{\psi}\tilde{\gamma}^\mu\partial_\mu\psi\\&-e^{3\alpha(\Phi)}m\bar{\psi}\psi \biggr),
\end{split}
\end{equation}
where $\psi$ is a Dirac spinor, with mass $m$, and $\gamma^\mu$ are the Dirac matrices in the local frame.
The field $\Phi$ is then written explicitly in terms of the matter fields solving its field equation, namely Eq.~\eqref{eqEF2}. Since for the model taken into consideration the solution to this equation is not analytical, it is necessary to solve it perturbatively. 

In this model one can perform two different expansions, one assuming that $\mu^2/\kappa^2T\ll 1$ and one considering $\kappa^2T/\mu^2\ll 1$. In Ref.~\citep{Flanagan:2003rb}, only the latter has been considered. This applies assuming that $\mu^4/R\gg R$ or equivalently $\mu^2\gg R\sim\kappa^2 T$. Hence, in this case one does not retrieve GR.
Using this expansion in order to solve the field equation of the scalar $\Phi$ yields
\begin{equation}
\label{solPhi}
\kappa\Phi=\kappa\Phi_\text{max}-\frac{3}{4\sqrt{2}}\mathcal{M}+\frac{1}{2\sqrt{2}}\mathcal{K}+O(\mathcal{K}^2,\mathcal{M}^2,\mathcal{M}\mathcal{K}),
\end{equation}
where $\kappa\Phi_\text{max}=\sqrt{\frac{3}{2}}\log\left(\frac{4}{3}\right)$ and $\mathcal{K}$ and $\mathcal{M}$ are 
dimensionless quantities defined as $\mathcal{K}=i\kappa^2\bar{\psi}\tilde{\gamma}^\mu\partial_\mu\psi/\mu^2$, 
$\mathcal{M}=\kappa^2m\bar{\psi}\psi/\mu^2$. Substituting Eq.~\eqref{solPhi} back into the action allows one to integrate out the scalar field $\Phi$. The action so obtained contains new matter interaction vertices each one characterized by the factor $1/m_* ^4$, where $m_*=\sqrt{\mu/\kappa}$. Since $\kappa^2=8\pi G$ and $\mu$ is assumed to be $\propto H_0$ to address the dark energy problem, it turns out that $m_*^4$ is roughly the geometric mean of the Planck and the Hubble scales if expressed in natural units, of order $10^{-3}\,{\rm eV}$. Therefore, Flanagan states that this model is in severe conflict with particle-physics experiments.

We note that these corrections are present at the tree level  in the Einstein frame. This point was the subject of the discussion in Ref.~\citep{Vollick:2003ic}. It was claimed there that the Jordan frame is the ``physical'' frame and that  these new matter interactions are absent in that frame (at tree level), since the matter-field equations do not present any correction terms. 
We chose here to revisit this discussion to address the issue of performing calculations in different conformal frames, which not only affects Palatini $f({\cal R})$ gravity, but more in general gravity theories with auxiliary fields. In the next section, we will look at the perturbed field equations in the two conformal frames, Jordan and Einstein, and we show that, as expected, both conformal frames can be used to do the calculation, but the change of frame affects the order in perturbation theory in which the corrections first appear.


\section{Matter fields in Palatini $f(\cR)$ gravity: a specific example}\label{Sec:MatterFields}

In this section we shall show that the matter-field equations in Palatini $\fR$ gravity are affected by corrections with respect to their GR counterpart in the local frame. 
We do not consider the $1/{\cal R}$ model as in Ref.~\citep{Flanagan:2003rb}, but instead we study the $\mathcal{R}^2$ model. Working with the latter, which has been used to describe inflationary scenarios, does not only simplify the calculations but also allows us to demonstrate that this issue is not strictly related to the $1/{\cal R}$ model. For completeness, we also report in Appendix~\ref{App:SecondModel} the analysis done for the specific theory studied in Ref.~\citep{Flanagan:2003rb} and reach the same conclusion.

\subsection{Generic considerations}\label{Sec:GenCons}

We wish to study the scalar-tensor representation of the theory $f(\mathcal{R})=\mathcal{R}+\beta\mathcal{R}^2$. First of all, we write explicitly the auxiliary scalar field as $\phi\equiv f'(\mathcal{R})=1+2\beta \mathcal{R}$. Solving Eq.~\eqref{algebraic}, one finds $\mathcal{R}=-\kappa^2 T$, and hence
\begin{equation}
\label{auxscalar1}
\phi=1-2\beta \kappa^2 T.
\end{equation}
Note that one could get the same result by rewriting $\mathcal{R}$ and $f(\mathcal{R})$ as functions of $\phi$ and solving Eq.~\eqref{eq:potential}.
Using now Eq.~\eqref{auxscalar1}, one can rewrite the potential $V(\phi)$ and its derivative $V'(\phi)$ in terms of the matter source,
\begin{equation}
V=\beta \kappa^4 T^2\,,\qquad V'=-\kappa^2 T\,.\label{potential1}
\end{equation}

Substituting the above relations into the modified Einstein equation~\eqref{eq:Jordan} yields
\begin{equation}
\label{eqSmunu1}
\begin{split}
G_{\mu\nu}& =\frac{\kappa^2 T_{\mu\nu}}{1-2\beta\kappa^2T}-\frac{1}{2}\frac{\beta\kappa^4g_{\mu\nu}T^2}{1-2\beta\kappa^2T}\\ &-3\beta^2\kappa^4\frac{2\nabla_\mu T\nabla_\nu T-g_{\mu\nu}\nabla^\lambda T\nabla_\lambda T}{(1-2\beta\kappa^2T)^2} \\&+2\beta\kappa^2\frac{g_{\mu\nu}\Box T-\nabla_\mu\nabla_\nu T}{1-2\beta\kappa^2T}.
\end{split}
\end{equation}

Since $\nabla^\mu G_{\mu\nu}=0$, the right-hand side of Eq.~\eqref{eqSmunu1} must be divergence-free. One can show this by employing the relation $(\square \nabla_\nu-\nabla_\nu \square)T=R_{\mu\nu}\nabla^\mu T$ and by rewriting the Ricci tensor $R_{\mu\nu}$ in terms of the matter fields only, by taking the trace of Eq.~\eqref{eqSmunu1}. Then, the right-hand side of Eq.~\eqref{eqSmunu1} vanishes identically provided $\nabla^\mu T_{\mu\nu}=0$.

For completeness, let us consider the same theory in the Einstein frame. In order to find an equivalent to Eq.~\eqref{auxscalar1} in this frame, we solve Eq.~\eqref{eqEF2}, finding
\begin{equation}
\label{auxscalar2}
\phi=\frac{\Delta-1}{4\beta\kappa^2 \tilde{T}},
\end{equation}
where we defined $\Delta =\sqrt{1+8\beta\kappa\tilde{T}}$.
Using now Eq.~\eqref{auxscalar2}, one can rewrite the potential $U(\phi)$ in terms of the matter source,
\begin{equation}
\label{eqU}
U=\frac{1+8\beta^2\kappa^4\tilde{T}^2-\Delta+4\beta\kappa^2\tilde{T}(2-\Delta)}{4\beta\kappa^2(\Delta-1)^2}.
\end{equation}

We can now write the Einstein modified equation~\eqref{eqEF2} using the relation just found, and the result is
\begin{equation}
\label{eqEF3}
\begin{split}
\tilde{G}_{\mu\nu}& =\kappa^2 \tilde{T}_{\mu\nu} 
-\kappa^2\tilde{g}_{\mu\nu}(\frac{1+8\beta^2\kappa^4\tilde{T}^2}{4\beta\kappa^2(1-\Delta)^2})\\& +\kappa^2 
\tilde{g}_{\mu\nu}(\frac{\Delta +4\beta\kappa^2\tilde{T}(\Delta-2)}{4\beta\kappa^2(1-\Delta)^2}).
\end{split}
\end{equation}
The modified Einstein equation derived in the Jordan frame is dynamically equivalent to the one derived in the Einstein frame and Eq.~\eqref{eqEF3} conformally transforms to Eq.~\eqref{eqSmunu1}.

So far, we performed an exact analysis of the Einstein modified equations. We are now going to study this model by performing an expansion in terms of the dimensionless parameter $\beta\kappa^2 T \ll 1$. This procedure is necessary to derive the matter-field equations in the Einstein frame in a simple form. Indeed, substituting in action~\eqref{actionE} the relation for the scalar field in terms of the matter source, i.e. Eq.~\eqref{auxscalar2}, yields a complicate expression. Moreover, the conformal transformation of the resulting equation back into the Jordan frame yields a cumbersome expression, unless the limit $\beta\kappa^2 T \ll 1$ is considered. Furthermore, this perturbative expansion also enlightens other properties of the field equations, as we shall see.

The modified Einstein equations in the Jordan frame, Eq.~\eqref{eqSmunu1}, reduce to
\begin{equation}
\label{eqSmunu1exp}
G_{\mu\nu}=\kappa^2 T_{\mu\nu}+S_{\mu\nu} \beta+{\cal O}(\beta^2),
\end{equation}
where $S_{\mu\nu}$ is the correction to the matter source defined as
\begin{equation}\label{eq:betaCorrection}
\begin{split}
S_{\mu\nu}&=2\kappa^4TT_{\mu\nu}-2\kappa^2\nabla_{\mu}\nabla_{\nu}T+2\kappa^2g_{\mu\nu}\square T\\
    &-\frac{ \kappa^4}{2}g_{\mu\nu}T^2.
\end{split}
\end{equation}
%
Let us now focus on the Einstein frame. Since $\tilde{T}_{\mu\nu}=\phi^{-1}T_{\mu\nu}$ and $\tilde{T}=\phi^{-2}T$, from Eq.~\eqref{auxscalar1} we have $T=\phi^{2}\tilde{T}=\tilde{T}+{\cal O}(\beta)$, and therefore
\begin{equation}
\phi=1-2\beta\kappa^2\tilde{T}+{\cal O}(\beta^2). \label{auxscalar1EF}
\end{equation}
We can use this relation to rewrite the potential $U(\phi)$ and its derivative $U'(\phi)$ as
\begin{eqnarray}
U&=&\frac{\beta\kappa^2\tilde{T}^2}{2(1-2\beta\kappa^2\tilde{T})^2}\,, \label{90}\\
U'&=&-\frac{1}{2}\frac{\tilde{T}}{(1-2\beta\kappa^2\tilde{T})^2}-\frac{\beta\kappa^2\tilde{T}^2}{(1-2\beta\kappa^2\tilde{T})^3}. \label{91}
\end{eqnarray}

Substituting these results into the modified Einstein equation~\eqref{eqEF1} and keeping terms up to ${\cal O}(\beta)$ yields
\begin{equation}
\label{92}
\tilde{G}_{\mu\nu}=\kappa^2\tilde{T}_{\mu\nu}-\frac{\kappa^2}{2}\beta\kappa^2\tilde{g}_{\mu\nu}\tilde{T}^2+{\cal O}(\beta^2),
\end{equation}
while Eq.~\eqref{eqEF2} is identically satisfied, as one would expect. Once again, the modified field equations derived in the Einstein frame are dynamically equivalent to those derived in the Jordan frame, namely Eq.~\eqref{eqSmunu1exp}.

The above considerations are valid for a generic matter stress-energy tensor. In the next section, we specialize to the specific case of a massless scalar field. This particular choice not only allows us to obtain more straightforward outcomes but also suffices to prove that matter corrections in Palatini $\fR$ gravity are a general result not only specific to the Dirac field, i.e. the case considered in Ref.~\citep{Flanagan:2003rb}. Nevertheless, in Appendix~\ref{app:Dirac}, we extend our result to the case of a Dirac field, reaching the same conclusion.

\subsection{Massless scalar field}\label{massless}

We consider the case of a massless scalar field described by the standard Klein-Gordon action,
\begin{equation}
\label{KG}
S_M=\frac{1}{2}\int d^4 x \sqrt{-g}\,g^{\mu\nu}\partial_\mu \psi \partial_\nu \psi,
\end{equation}
so that $T_{\mu\nu}=\frac{1}{2}g_{\mu\nu}\partial^\lambda\psi\partial_\lambda\psi-\partial_\mu\psi\partial_\nu\psi$ and $T=\partial^\lambda\psi\partial_\lambda\psi$. 
Part of the debate regarding the analysis of Ref.~\citep{Flanagan:2003rb} was related to whether corrections to the behavior of matter were specific to the Einstein frame. The Jordan and Einstein frames are equivalent and this has already been shown in Sec.~\ref{Sec:ScalTens} in the formulation in which the auxiliary field $\phi$ is present. A subtlety can arise when one decides to eliminate the auxiliary field.   One can always do so at the level of the field equations in both frames and obtain dynamically equivalent equations, as we show below. However, attempting to use the field equations to eliminate $\phi$ from the action in the Jordan frame will not yield the correct field equations, as the Jordan frame action~\eqref{actionST} contains derivatives of $\phi$ (despite the fact that $\phi$ can be algebraically determined from the field equations). Hence, discussing the equivalence of the two frames at the level of this action, after attempting to eliminate $\phi$ is misleading.
In Appendix~\ref{App:model}, we present a simple Lagrangian model to elucidate this point. Below we demonstrate it explicitly for our model.

Let us consider the modified Einstein equation in the Jordan frame, namely Eq.~\eqref{eqSmunu1}, and let us substitute the explicit definition of the stress-energy tensor directly at the level of the equation. The resulting equation is
\begin{equation}
\label{87}
\begin{split}
G_{\mu\nu}&=  \frac{1}{2}\kappa^2 g_{\mu\nu}\partial^\lambda\psi\partial_\lambda\psi-\kappa^2\partial_\mu\psi\partial_\nu\psi\\ &+\frac{1}{2}\beta\kappa^4g_{\mu\nu}\partial^\lambda\psi\partial_\lambda\psi\partial^\sigma\psi\partial_\sigma\psi\\&-2\beta\kappa^4\partial^\lambda\psi\partial_\lambda\psi\partial_\mu\psi\partial_\nu\psi-4\beta\kappa^2\nabla_\mu\partial_\lambda\psi\nabla_\nu\partial^\lambda\psi\\
            &-4\beta\kappa^2\partial_\lambda\psi\nabla_{(\mu}\nabla_{\nu)}\partial^\lambda\psi \\
            & +4\beta\kappa^2g_{\mu\nu}\nabla_\lambda\partial_\sigma\psi\nabla^\lambda\partial^\sigma\psi \\
            &+4\beta\kappa^2g_{\mu\nu}\partial_\sigma\psi\,\square\partial^\sigma\psi+{\cal O}(\beta^2).
\end{split}
\end{equation}

However, integrating out the scalar field at the level of the action, using the algebraic relation~\eqref{auxscalar1}, and varying the action with respect to the metric and the matter field yields a different field equation, respectively,
\begin{equation}
\label{wrong1}
\begin{split}
G_{\mu\nu}&=  \frac{1}{2}\kappa^2g_{\mu\nu}\partial_\lambda\psi\partial^\lambda\psi-\kappa^2\partial_\mu\psi\partial_\nu\psi\\&+2\beta\kappa^2R\,\partial_\mu\psi\partial_\nu\psi -2\beta\kappa^2\nabla_\mu\nabla_\lambda\partial_\nu\psi\partial^\lambda\psi\\&-2\beta\kappa^2\nabla_\nu\nabla_\lambda\partial_\mu\psi \partial^\lambda\psi+2\beta\kappa^2G_{\mu\nu}\partial_\lambda\psi\partial^\lambda\psi\\&+2\beta\kappa^4\partial_\lambda\psi\partial^\lambda\psi\partial_\mu\psi\partial_\nu\psi  -4\beta\kappa^2\nabla_\lambda\partial_\nu\psi\nabla^\lambda\partial_\mu\psi\\&-\frac{1}{2}\beta\kappa^4g_{\mu\nu}\partial_\lambda\psi\partial^\lambda\psi\partial_\sigma\psi\partial^\sigma\psi\\&+4\beta\kappa^2g_{\mu\nu}\partial^\lambda\psi\square\partial_\lambda\psi \\&+4\beta\kappa^2g_{\mu\nu}\nabla_\lambda\partial_\sigma\psi\nabla^\lambda\partial^\sigma\psi+{\cal O}(\beta^2)
\end{split}
\end{equation}
and
\begin{equation}
\label{wrong2}
\begin{split}
& \Box\psi-2\beta R\Box\psi-2\beta\nabla_\lambda\psi\nabla^\lambda R-2\beta\kappa^2\nabla_\lambda\psi\nabla^\lambda\psi \Box\psi\\
& -4\beta\kappa^2\nabla^\lambda\psi\nabla^\sigma\psi\nabla_\lambda\nabla_\sigma\psi+{\cal O}(\beta^2)=0.
\end{split}
\end{equation}
Clearly, Eq.~\eqref{wrong1} is not equivalent to Eq.~\eqref{87}. As we previously emphasized, substituting $\phi$ in terms of the matter field $\psi$ in order to write the action explicitly in terms of the latter is not appropriate, and, as we showed, leads to incorrect field equations. We stress that we do not expect the set of Eqs.~\eqref{wrong1} and \eqref{wrong2} to give the same dynamics as Eq.~\eqref{87} coupled with its respective matter-field equation. This would be hard to show more explicitly, but, as we pointed out, Eqs.~\eqref{wrong1} and \eqref{wrong2} describe a different theory. We refer to the simple model in Appendix~\ref{App:model} to clarify this aspect.\footnote{Note that, in the context of effective field theory where operators above a certain mass dimension or order in derivatives could be neglected, it might be possible to reconcile Eq.~\eqref{87} and Eq.~\eqref{wrong1}, via field redefinitions as discussed in Refs.~\cite{Bloomfield:2011np,Solomon:2017nlh}.} A consistent procedure is to derive the matter-field equation in the Einstein frame, where we can substitute the expression of $\phi$ in terms of $\psi$ directly in the action, since the auxiliary scalar field has no kinetic term in the action, and then transform the equation back into the Jordan frame. This allows us to use the action in the Einstein frame to draw an initial conclusion about the theory. 

Let us then consider our model in the Einstein frame. The Klein-Gordon action~\eqref{KG} can be written as
\begin{equation}
\label{93}
\tilde{S}_M=\frac{1}{2}\int d^4x\sqrt{-\tilde{g}}\phi^{-1}\tilde{g}^{\mu\nu}\partial_\mu\psi\partial_\nu\psi,
\end{equation}
so that $\tilde{T}_{\mu\nu}=\frac{1}{2}\tilde{g}_{\mu\nu}\phi^{-1}\partial^\lambda\psi\partial_\lambda\psi-\phi^{-1}\partial_\mu\psi\partial_\nu\psi$ and $\tilde{T}=\phi^{-1}\tilde{g}^{\mu\nu}\partial_\mu\psi\partial_\nu\psi$. Thus, Eq.~\eqref{92} can be written explicitly in terms of the matter field $\psi$ as
\begin{equation}
\label{94}
\begin{split}
\tilde{G}_{\mu\nu}&=\frac{\kappa^2}{2}\tilde{g}_{\mu\nu}\partial_\sigma\psi\partial^\sigma\psi-\kappa^2\partial_\mu\psi\partial_\nu\psi\\
&+\frac{1}{2}\tilde{g}_{\mu\nu}\beta\kappa^4\partial_\sigma\psi\partial^\sigma\psi\partial_\lambda\psi\partial^\lambda\psi \\ &-2\beta\kappa^4\partial_\sigma\psi\partial^\sigma\psi\partial_\mu\psi\partial_\nu\psi+{\cal O}(\beta^2).
\end{split}
\end{equation}
As mentioned before, in the Einstein frame the auxiliary field $\phi$ has no kinetic term in the action. Thus, in this frame we can substitute the solution~\eqref{auxscalar1EF} for $\phi$ in terms of $\psi$ directly into the action. Varying with respect to the metric and the matter field, we obtain the correct field equations. Indeed, action~\eqref{actionE} can be written explicitly in terms of the matter field $\psi$ as
\begin{equation}
\label{95}
\begin{split}
&\tilde{S}=\int d^4x \sqrt{-\tilde{g}}\biggl(\frac{\tilde{R}}{2\kappa^2} +\frac{\partial_\sigma\psi\partial^\sigma\psi}{2(1-2\beta\kappa^2\partial_\sigma\psi\partial^\sigma\psi)}\\&-\frac{\beta\kappa^2\partial_\sigma\psi\partial^\sigma\psi\partial_\lambda\psi\partial^\lambda\psi}{2(1-4\beta\kappa^2\partial_\sigma\psi\partial^\sigma\psi+4\beta^2\kappa^4\partial_\sigma\psi\partial^\sigma\psi\partial_\lambda\psi\partial^\lambda\psi)} \biggr),
\end{split}
\end{equation}
varying the action with respect to the metric and keeping terms up to ${\cal O}(\beta)$, we retrieve exactly the field equation~\eqref{94}. Finally, varying the action with respect to $\psi$ instead, we obtain the modified matter-field equation,
\begin{equation}\label{96}
\begin{split}
&\tilde{\square}\psi+2\beta\kappa^2\,\tilde{\square}\psi\,\partial_\lambda\psi\partial^\lambda\psi\\&+4\beta\kappa^2\,\partial^\lambda\psi\partial^\sigma\psi\,\tilde{\nabla}_\sigma\partial_\lambda\psi+{\cal O}(\beta^2)=0.
\end{split}
\end{equation}
As expected, in the Einstein frame the standard Klein-Gordon equation is modified by ${\cal O}(\beta)$ terms due to the nonminimal coupling in the Einstein frame. These corrections are evident in Eq.~\eqref{96}, but can already be identified as modifications to the standard kinetic term in the third term of action~\eqref{95} (note that at order $\beta$ the numerator is simply a constant). These terms are suppressed by a mass scale proportional to $\mathcal{M}\propto1/(\beta^{1/4}\sqrt{\kappa})$, instead of the usual Planck mass scale, that is $M_{\rm Planck}\propto 1/\kappa$. Hence, as discussed in more detail below, to guarantee that these corrections do not appear at a larger scale one must severely constrain the parameter $\beta$.

We now show that these terms persist in the Jordan frame field equations as well, i.e. the frame where we want to verify if one locally retrieves the correct behavior of the matter fields predicted by the Standard Model. Let us perform a conformal transformation on the field equations~\eqref{94} and~\eqref{96}. It is straightforward to show that Eq.~\eqref{94} becomes exactly the same field equation previously found in the Jordan frame, i.e. Eq.~\eqref{87} and not the wrong one, namely Eq.~\eqref{wrong1}. Finally, a conformal transformation of Eq.~\eqref{96} yields
\begin{equation}\label{98}
\begin{split}
&\square\psi(1+4\beta\kappa^2\partial_\lambda\psi\partial^\lambda\psi)+{\cal O}(\beta^2)\\&=\phi^{-2}\square\psi+{\cal O}(\beta^2)=0.
\end{split}
\end{equation}
Since we are working perturbatively in the regime $\beta\kappa T \ll 1$, this equation implies that $\square\psi=0$.
However, the matter field is on-shell only at tree level, i.e. at first order in perturbation theory. Indeed, performing an expansion in the field around a flat background, that is $g_{\mu\nu}=\eta_{\mu\nu}+\epsilon\delta g^1_{\mu\nu}+\cdots$ and $\psi=\epsilon \delta\psi^1+\cdots$~\footnote{We have assumed that $\psi^0=\text{const}=0$ for simplicity.}, the modified Einstein equation and the matter-field equation at first order, respectively, are
\begin{equation}\label{eq:FirstOrderGravity}
\begin{split}
& O_\text{g}(\delta g^1)\equiv 2\partial^\lambda\partial_{(\mu}\delta g^1_{\nu )\lambda}-\partial_\mu\partial_\nu\delta g^1-\eta_{\lambda\sigma}\partial^\lambda\partial^\sigma\delta g^1_{\mu\nu} \\
& -\eta_{\mu\nu}\partial^\lambda\partial^\sigma\delta g^1_{\lambda\sigma}+\eta_{\mu\nu}\eta_{\lambda\sigma}\partial^\lambda\partial^\sigma\delta g^1=0,
\end{split}
\end{equation} 
\begin{equation}\label{eq:FirstOrderMatter}
 O_{\text{m}}(\delta\psi^1)\equiv\eta_{\mu\nu}\partial^\mu\partial^\nu \delta\psi^1=0,
\end{equation}
where we identified $O_\text{g}$ and $O_\text{m}$ as the differential operators acting on $\delta g^1$ and $\delta\psi^1$, respectively. Equation~\eqref{eq:FirstOrderMatter} is the field equation for a Klein-Gordon massless field in the local frame predicted by the Standard Model. The fact that, in the Jordan frame, no modifications appear at tree level was the source of the confusion in the debate on the results of Ref.~\citep{Flanagan:2003rb}. However, these corrections do appear at higher order in perturbation theory, as we now show.

Indeed, at higher order the two perturbed equations have, respectively, the following structure:
\begin{equation}\label{eq:NextOrderGravity}
O_\text{g}(\delta g^n)=S_{\text{g}}(\delta g^0,\ldots\,,\delta g^{n-1},\delta\psi^1,\ldots\,,\delta\psi^{n-1}),
\end{equation}
\begin{equation}\label{eq:NextOrderMatter}
O_{\text{m}}(\delta\psi^n)=S_{\text{m}}(\delta g^0,\ldots\,,\delta g^{n-1},\delta\psi^1,\ldots\,,\delta\psi^{n-1}),
\end{equation}
where $O_\text{g}$ and $O_\text{m}$ are the same differential operators defined in Eqs.~\eqref{eq:FirstOrderGravity} and \eqref{eq:FirstOrderMatter}, $n$ is the order of perturbation considered, and the terms in the right-hand side act as a source for $\delta g^n$ and $\delta\psi^n$; e.g. $\delta g^2$ is sourced by $\delta\psi^1$ and in turn sources $\delta\psi^3$. This mixing mechanism introduces matter corrections to the matter-field equation that can be big and therefore are in conflict with particle physics. Indeed, some of the terms that source $\delta g^2$ and percolate to the matter sector already at third order are not Planck suppressed. These corrections come from the gravity sector and are already manifest in the last four terms on the right-hand side of Eq.~\eqref{87}. They introduce modifications to the perturbation of the metric $\delta g$ with a mass scale $\mathcal{M}\propto1/(\beta^{1/4}\sqrt{\kappa})$, which in turn will source higher-order corrections in perturbation theory in $\delta\psi$. We do not report here the resulting lengthy equations, we instead refer the reader to a \textit{Mathematica} notebook including the perturbed equations which is publicly available at Ref.~\citep{Ventagli2023}. For example, this mechanism introduces a correction to the d'Alembertian of $\delta\psi^3$ proportional to $(\beta\kappa^2\nabla\delta\psi^1\nabla\delta\psi^1)\nabla^2\delta\psi^1$, a term that lacks an additional $\kappa^2$ factor that would guarantee that this is a correction at the Planck scale. To avoid the appearance of this term at a larger scale, which would invalidate the whole perturbation scheme, one must then severely constrain the parameter $\beta$ to guarantee that the theory is in agreement with current particle observations. Providing precise constraints on $\beta$ goes beyond the scope of our analysis, but we can give a simple order-of-magnitude estimate as follows. Schematically, the problematic terms introduce correction factors of the form
\begin{equation}
    (1+\beta\kappa^2 T)\,,
\end{equation}
and the standard-model result is obtained when $\beta\to0$. The second term in the above parenthesis should be much smaller than unity, both to be consistent with our perturbative expansion and also because otherwise it would introduce ${\cal O}(1)$ corrections incompatible with particle-physics experiments. Thus, we can estimate that if
\begin{equation}
    \beta\kappa^2 T\sim 1  \,,
\end{equation}
or higher, the corrections will drastically modify the usual dynamics. The trace $T$ is dimensionally an energy density and, working in $c=1$ units, $T\sim m^4/\hbar^3$, where $m$ is the mass scale of a given standard-model particle. Since $\kappa^2=\hbar/M_{\rm Planck}^2$, we get
\begin{equation}
    \beta \sim \frac{M_{\rm Planck}^2}{m^4}\hbar^2 \sim 0.2\left(\frac{{\rm GeV}}{m}\right)^4 {\rm km}^2\,, \label{scalebeta}
\end{equation}
where we have normalized $m$ to a typical standard-model scale.
This value of $\beta$ corresponds to a very low energy scale
\begin{equation}
    \frac{\hbar}{\beta^{1/2}}\sim 4\times 10^{-10}\left(\frac{m}{{\rm GeV}}\right)^2 {\rm eV}\,.
\end{equation}
%
In other words, already at sub-eV scales the corrections to the usual standard-model dynamics are dominant and incompatible with particle-physics experiments.

To restore compatibility with experiments, the only possibility is to push the energy scale of these corrections at least at the ${\cal O}(10\,{\rm TeV})$ level.
This requires decreasing $\beta$ by \emph{forty orders of magnitude} relative to the scale in Eq.~\eqref{scalebeta}, in which case the correction would not have any impact on the macroscopic dynamics relevant for gravitational theories (e.g., compact objects or cosmological evolution).

Thus, we have demonstrated that Palatini $\fR$ gravity is incompatible with the Standard Model, unless the parameters of the theory are carefully fine-tuned to a level in which they are phenomenologically irrelevant.

Note that the problematic terms responsible for the appearance of new matter corrections are already manifest in Eq.~\eqref{eqfieldMod}, or equivalently in Eq.~\eqref{eq:Jordan}, and they correspond to those terms proportional to derivatives of the stress-energy tensor and of its trace.

Although we anticipate that the results just obtained are valid in general, for completeness, in Appendix~\ref{app:Dirac} we extend our analysis to the case of a Dirac field which was also considered in Ref.~\citep{Flanagan:2003rb}, finding agreeing conclusion.
%

 
\section{Gravity with auxiliary fields}\label{Sec:AuxiliaryFields}

The mechanism affecting the field equations of Palatini $\fR$ gravity is a more general problem regarding theories with auxiliary fields.

As previously mentioned, this generic class of theories has been studied in Ref.~\citep{Pani:2013qfa}, where a general parametrization has been proposed such that, to the next-to-leading order in derivatives of the matter fields, this class of theories can be described with only two parameters. However, the presence of higher-order derivatives of the matter fields in the field equations causes the metric to be overly sensitive to abrupt discontinuities in the matter energy density leading to spacetime singularities, e.g. on the surface of stars. This phenomenon puts severe constraints on the parameters of the theory.

In this section we show that the same terms causing spacetime singularities, studied in Ref.~\citep{Pani:2013qfa}, can also lead to conflicts with the Standard Model.

The modified Einstein field equations for generic gravity theories with auxiliary fields up to fourth order in derivative is~\citep{Pani:2013qfa}
\begin{equation}
\begin{split}\label{eq:AuxField}
    G_{\mu\nu} & =T_{\mu\nu}-\Lambda g_{\mu\nu}-\beta_1\Lambda g_{\mu\nu}T\\
    &+\frac{1}{4}(1-2\beta_1\Lambda)(\beta_1-\beta_4)g_{\mu\nu}T^2\\
    & +[\beta_4(1-2\,\beta_1\Lambda)-\beta_1]T\,T_{\mu\nu}\\
    &+\frac{1}{2}\beta_4 g_{\mu\nu}T_{\lambda \kappa}T^{\lambda \kappa}-2\,\beta_4 {T^\kappa}_\mu T_{\kappa\nu}\\
    &+\beta_1\nabla_\mu\nabla_\nu T-\beta_1 g_{\mu\nu}\DAlembert{T} -\beta_4\DAlembert{T_{\mu\nu}}\\
    &+2\,\beta_4 \nabla^\kappa\nabla_{(\mu}T_{\nu)\kappa}+\cdots,
\end{split}
\end{equation}
where $\beta_1$ and $\beta_4$ are coefficients with appropriate dimensions parametrizing the theory and where $\kappa^2=1$. Known theories with auxiliary fields are indeed described by Eq.~\eqref{eq:AuxField}. Palatini $\fR$  gravity corresponds to $\beta_4=0$ with $\Lambda$ and $\beta_1$ depending on the specific model taken into considerations. For example, the quadratic model we considered in Sec.~\ref{Sec:MatterFields} is retrieved by taking $\Lambda=0$ and $\beta_1=-2\beta$, whereas the model studied in Ref.~\citep{Flanagan:2003rb}, i.e. $\fR={\cal R}-\mu^4/{\cal R}$, performing the expansion $\kappa^2 T\gg \mu^2$ corresponds to $\Lambda=0$ and $\beta_1=\mu^4/T^3$, after the rescaling $\kappa^2=1$. \footnote{In Ref.~\citep{Flanagan:2003rb}, the opposite expansion is performed, that is $\mu^2\gg\kappa^2 T$. This is more interesting from a late-time cosmology perspective. In this case, however, a mapping to Eq.~\eqref{eq:AuxField} is not possible, as the latter is obtained as a gradient expansion and by assigning $T$ to be equivalent to 2 derivatives for the purposes of that expansion~\citep{Pani:2013qfa}. This turns out to be incompatible with the $\mu^2\gg\kappa^2 T$ expansion.}

Let us now perform an expansion in the fields around a flat background for the case when matter is described by a massless scalar field. At first order, we retrieve Eqs.~\eqref{eq:FirstOrderGravity} and~\eqref{eq:FirstOrderMatter}, with the addition of a cosmological constant term to the modified Einstein equation. Nonetheless, at higher order we recover the same behavior we highlighted for the specific case of Palatini $\fR$ gravity; that is the field equations follow the structure of Eqs.~\eqref{eq:NextOrderGravity} and \eqref{eq:NextOrderMatter}. For the complete set of perturbed equations we refer to the publicly available \textit{Mathematica} notebook at Ref.~\citep{Ventagli2023}. For example, at second order $\delta g^2$ is sourced by terms proportional to $\beta_1\nabla^2\delta\psi^1\nabla^2\delta\psi^1$ and $\beta_4\nabla^2\delta\psi^1\nabla^2\delta\psi^1$. In turn, $\delta g^2$ appears at third order in the matter-field equations. We thus have $\delta\psi^3$ sourced by terms proportional to $(\beta_1\nabla\delta\psi^1\nabla\delta\psi^1)\nabla^2\delta\psi^1$ and $(\beta_4\nabla\delta\psi^1\nabla\delta\psi^1)\nabla^2\delta\psi^1$, which are contributions that are not Planckian suppressed. As for the specific case of Palatini $\fR$ gravity, the terms responsible for these new matter-field corrections are those proportional to derivatives of the stress-energy tensor and its trace in Eq.~\eqref{eq:AuxField}. Guaranteeing that the theory is in agreement with particle-physics observations would require tight constraints on the parameters describing the theory, i.e. $\beta_1$ and $\beta_4$. It then seems challenging to construct a theory with auxiliary fields that can introduce effects at large scales, while being in agreement with the Standard Model.


\section{Discussion}\label{Sec:Discussion}

We have shown that Palatini $\fR$ gravity introduces new matter-field interactions incompatible with current particle physics observation unless one severely constraints the parameters of the theory. This prevents Palatini $\fR$ gravity to produce significant phenomenology at large scales, confirming and elucidating the results obtained in Ref.~\cite{Flanagan:2003rb}. 

Palatini $\fR$ gravity can be recast as a particular scalar-tensor theory, where the scalar can be algebraically determined by the field equations. Hence it carries no dynamics and is an \textit{auxiliary} field. In the Einstein frame, where this field couples directly to matter but only couples minimally to gravity, one can eliminate it at the level of the action. This changes the matter action and hence the field equations of the matter fields, as shown in Ref.~\cite{Flanagan:2003rb} and verified here.

In the Jordan frame, where the scalar couples nonminimally to gravity but does not couple to matter, one is tempted to think that eliminating it cannot introduce modifications to the matter-field equations. Indeed, this has been the subject of debate \cite{Vollick:2003ic,Flanagan:2004bz}. Considering the effects on matter in the Jordan frame is subtle, because one cannot formally eliminate the auxiliary scalar at the level of the action without affecting the dynamics. This is because the action contains derivatives of the scalar. We have performed a perturbative analysis of the field equations instead, where the scalar can be consistently eliminated. The matter-field equations are retrieved unaffected at tree level, but at higher order new matter interactions do percolate from the gravity into the matter sector, in agreement with the Einstein frame analysis. We explicitly showed this mechanism for the specific case of a quadratic $\fR$ model, i.e. $\fR=\mathcal{R}+\beta\mathcal{R}^2$, considering a massless scalar field. In Appendix~\ref{app:Dirac}, we explore the case of a Dirac field, finding agreeing results. We also performed a similar analysis for the case considered in Ref.~\cite{Flanagan:2003rb}, which we report in Appendix~\ref{App:SecondModel}.

Our analysis also clearly demonstrated that the corrections to the matter-field equations are not suppressed by the Plank scale, as would happen for correction to the Standard Model coming from gravitons. Instead, the additional interaction terms contain the scale of the coupling that controls the correction to the Einstein-Hilbert actions, {\em e.g.}~$\beta$ in the case of $\fR=\mathcal{R}+\beta\mathcal{R}^2$. Hence, if that energy scale is low in order to produce deviation from GR at low energies, it is bound to produce sizable deviations from the Standard Model as well.

We have shown that this shortcoming is not specific to Palatini $\fR$ gravity but instead it is a general feature for theories with auxiliary fields. Interestingly, the same terms responsible for the appearance of new matter-field interactions were shown to introduce spacetime singularities, e.g. on the surface of stars, in Ref.~\citep{Pani:2013qfa}. Requiring that such theories are in agreement with the Standard Model of particle physics would severely constrain the parameters of these models, preventing them from producing significant phenomenology at large scales.
It is worth noting that similar conclusions were reached in Refs.~\cite{Latorre:2017uve,Delhom:2019wir,BeltranJimenez:2021oaq} in the context of a wide class of metric-affine theories of gravity, also known as Ricci-Based gravity, which includes $\fR$ gravity. In this class of theories the metric and the connection are two independent fields, while the latter is also allowed to contain torsion. The nonmetricity of the theory produces nontrivial effective interactions which can be used to impose tight constraints on the model parameters, hinting that this might be a generic feature of theories where the dynamical metric is built through a field redefinition of some auxiliary fields.

\begin{acknowledgments}
We are indebted to Eanna Flanagan for interesting comments on the draft.
P.P. acknowledges financial support provided under the European
Union's H2020 ERC, Starting Grant Agreement No.~DarkGRA--757480 and under the MIUR PRIN program. 
G.V. and T.P.S. acknowledge financial support from EPSRC Grant No.~EP/W52251X/1. T.P.S. acknowledges partial support from
the STFC Consolidated Grants No.~ST/T000732/1 and
No.~ST/V005596/1.
The authors acknowledge networking support by the COST Action CA16104.
\end{acknowledgments}
%

\appendix
\section{A LAGRANGIAN TOY MODEL WITH A NONDYNAMICAL DEGREE OF FREEDOM}\label{App:model}

We propose here a simple one-dimensional Lagrangian model with 2 dynamical degrees of freedom, $q=q(t)$ and $u=u(t)$, and a nondynamical one, $h=h(t)$, which is algebraically related to $q$ and $u$. Our goal is to show that if the Lagrangian contains a kinetic term of the auxiliary degree of freedom, one cannot substitute the explicit expression of $h(q,u)$ directly in the Lagrangian, since it would lead to the wrong equations of motion.

Let us consider the following model:
\begin{equation}
\label{B1}
\mathcal{L}=\frac{1}{2}\dot q^2+\frac{1}{2}q^2+\frac{1}{2}\dot h^2+qh+\dot q \dot h+q\dot u^2.
\end{equation}
The Euler-Lagrange equations give
\begin{eqnarray}
 \frac{d}{dt}\frac{\partial \mathcal{L}}{\partial\dot q}-\frac{\partial \mathcal{L}}{\partial q}&=&\ddot q+\ddot h-q-h-\dot u^2 =0 \,, \label{B2} \\
 \frac{d}{dt}\frac{\partial \mathcal{L}}{\partial\dot h}-\frac{\partial \mathcal{L}}{\partial h}&=&\ddot h+\ddot q-q=0\,,\label{B3}\\
 \frac{d}{dt}\frac{\partial \mathcal{L}}{\partial\dot u}-\frac{\partial \mathcal{L}}{\partial u}&=&2\dot q \dot u+2q\ddot u=0 \,.\label{B4}
\end{eqnarray}
From Eq.~\eqref{B2}, one finds
\begin{equation}
\label{B5}
\ddot q=-\ddot h+q+h+\dot u^2,
\end{equation}
and substituting this back into Eq.~\eqref{B3} yields an algebraic relation between $h$ and $u$, namely 
\begin{equation}
 h=-\dot u^2\,. \label{B7}
\end{equation}
Using this equation to rewrite Eqs.~\eqref{B2}-\eqref{B4} yields
\begin{eqnarray}
\ddot q-q-2\ddot u^2-2\dot u\dddot u&=&0\,, \label{B6}\\
2\dot q \dot u+2q\ddot u&=&0\,.  \label{B8}
\end{eqnarray}
It is clear that $q$ and $u$ are dynamical, whereas $h$ is just an auxiliary degree of freedom.

Let us now show that substituting the relation~\eqref{B7} directly in the Lagrangian~\eqref{B1} leads to a different theory; i.e.\ the equations of motion derived from the new Lagrangian are not equivalent to Eqs.~\eqref{B6} and \eqref{B8}.
We first substitute the algebraic relation between $h$ and $u$ in Eq.~\eqref{B1}. The new Lagrangian $\mathcal{L}'$ so obtained is
\begin{equation}
\label{B9}
\mathcal{L}'=\frac{1}{2}\dot q^2+\frac{1}{2}q^2+2\dot u^2\ddot u^2-2\dot q\dot u \ddot u.
\end{equation} 
Note that $\mathcal{L}'$ does not contain the function $u$, but only its first and second derivatives. We then proceed to define a new function $v\equiv \dot u$, so that $\mathcal{L}'$ can be written as
\begin{equation}
\label{B10}
\mathcal{L}=\frac{1}{2}\dot q^2+\frac{1}{2}q^2+2v^2\dot v^2-2\dot q v \dot v.
\end{equation}
The equations of motion for this Lagrangian are
\begin{eqnarray}
\ddot q-q-2\dot v^2-2v\ddot v&=&0\,, \label{B11} \\
4v\dot v^2+4v^2 \ddot v-2\ddot q v&=&0\,,\label{B12} \\
v- \dot u&=&0\,. \label{B13}
\end{eqnarray}
We now use Eq.~\eqref{B13} to simplify the system of equations and write it only in terms of $q$ and $u$ as
\begin{eqnarray}
\ddot q-q-2\ddot u^2-2\dot u\dddot u&=&0, \label{B14} \\
4\dot u\ddot u^2+4\dot u^2 \dddot u-2\ddot q \dot u&=&0. \label{B15}
\end{eqnarray}
Note that, while Eq.~\eqref{B14} is equivalent to Eq.~\eqref{B6}, Eq.~\eqref{B15} is different from the one previously retrieved, namely Eq.~\eqref{B8}.
Therefore, the two Lagrangians~\eqref{B1} and~\eqref{B10} are not dynamically equivalent.

The problem with this derivation can be traced back to the fact that, in the original Lagrangian~\eqref{B1}, the nondynamical nature of $h$ is not manifest. Indeed, only after having derived and manipulated the equations of motion does one find that $h$ is an auxiliary degree of freedom. Forcing this information \textit{a priori} into the Lagrangian, 1 degree of freedom is lost, modifying the original theory.

The model described by Eq.~\eqref{B1} mimics, in a very simplified way, the action in the Jordan frame for a scalar-tensor version of Palatini $\fR$ theory. In the Lagrangian~\eqref{B1}, $q$, $h$, and $u$ play the role of the metric, the scalar and the matter field, respectively. 

Finally, we note that substituting the explicit expression, obtained from the equations of motion, of a nondynamical degree of freedom in terms of a dynamical one directly into the Lagrangian does not always lead to a different theory. As an example, let us consider the Lagrangian
\begin{equation}
\label{B16}
\mathcal{L}=\frac{1}{2}\dot q^2-\frac{1}{2}q^2+\frac{1}{2}\dot h^2+qh+\dot q \dot h\,.
\end{equation}
In this case the Euler-Lagrange equations yield $h=2q$, and, upon substitution in $\mathcal{L}$, it is easy to see that the equation for $q$ derived from the new Lagrangian is equivalent to the one derived from Eq.~\eqref{B16}; i.e. the two theories are dynamically equivalent.
Therefore, the aforementioned problem in deriving the equations of motion seems to arise from the fact that the auxiliary field $h$ is algebraically related to the {\it derivatives} of another dynamical field. This, upon direct substitution in the Lagrangian, introduced higher-order derivatives for the dynamical field, which affect the Euler-Lagrange equations.


\section{DIRAC FIELD}\label{app:Dirac}
%
We now extend the results obtained for the $f({\cal R})={\cal R}+\beta {\cal R}^2$ model to the case of a Dirac field.
In the Jordan frame the Dirac matter action is\footnote{We stress that in action~\eqref{113} we used the partial derivative $\partial_\mu\psi$ instead of the covariant derivative usually defined for spinors, that is $D_\mu\psi=\nabla_\mu\psi-\Gamma_\mu\psi=\partial_\mu\psi-\Gamma_\mu\psi$. The reason for this substitution is to be found in one of the initial assumptions of Palatini $f(\cR)$ theory. Indeed, we imposed that the independent connection is not present in the matter action, and for a spinor this implies that there is no torsion. Therefore, in our case, the covariant derivative coincides with the partial derivative, i.e. $D_\mu\psi=\partial_\mu\psi$.}
\begin{equation}
\label{113}
S_M=\int d^4x \sqrt{-g} \bar{\psi}(i \gamma^\mu \partial_\mu - m)\psi.
\end{equation}
Deriving the stress-energy tensor and its trace from action~\eqref{113} leads to $T_{\mu\nu}=g_{\mu\nu}\bar{\psi}(i \slashed{\partial} - m)\psi-\bar{\psi}i\gamma_\mu\partial_\nu\psi$ and $T=3\bar{\psi}i\slashed{\partial}\psi-4m\bar{\psi}\psi$.

We now focus on the field equations. We shall derive only the matter-field equation, since we have already showed in Sec.~\ref{Sec:GenCons} that the modified Einstein equations in the two conformal frames are dynamically equivalent, keeping the stress-energy tensor implicit. 
As in the previous case, we rewrite the theory in the Einstein frame,
\begin{equation}
\label{117}
\tilde{S}_M=\int d^4x \sqrt{-\tilde{g}}\bar{\psi}(i\phi^{-3/2}\tilde{\gamma}^\mu\partial_\mu-\phi^{-2}m)\psi,
\end{equation}
where $\tilde{\gamma}^\mu$ are the Dirac matrices\footnote{The Dirac matrices associated with the metric $g_{\mu\nu}$ satisfy $\left\{ \gamma^\mu ,\gamma^\nu \right\}=2g^{\mu\nu}$, whereas the Dirac matrices in the Einstein frame satisfy $\left\{ \tilde{\gamma}^\mu ,\tilde{\gamma}^\nu \right\}=2\tilde{g}^{\mu\nu}$. Thus, since $\tilde{g}^{\mu\nu}=\phi^{-1} g^{\mu\nu}$, the relation between the two Dirac matrices is $\gamma^\mu=\phi^{1/2}\tilde{\gamma}^\mu$.} in the frame described by the metric $\tilde{g}_{\mu\nu}$.  Then, substituting the solution for $\phi$ in the total action~\eqref{actionE} and keeping terms up to ${\cal O}(\beta)$ yields
\begin{equation}
\label{117b}
\begin{split}
\tilde{S} & =\int d^4x \sqrt{-\tilde{g}}\Bigg[ \frac{\tilde{R}}{2\kappa^2}-\frac{1}{2}\beta\kappa^2T^2+\bar{\psi}(i\tilde{\gamma}^\mu\partial_\mu\\&+3i\beta\kappa^2T\tilde{\gamma}^\mu\partial_\mu-m-4\beta\kappa^2Tm)\psi+{\cal O}(\beta^2)\Bigg].
\end{split}
\end{equation}
Substituting the expression of $T$ and varying the action with respect to $\bar{\psi}$ gives
\begin{equation}
\label{118}
\begin{split}
&i\tilde{\gamma}^\mu\partial_\mu\psi-m\psi+\beta\kappa^2[ -9(\bar{\psi}\tilde{\gamma}^\mu\partial_\mu\psi)(\tilde{\gamma}^\mu\partial_\mu\psi)\\&-12i(m\bar{\psi}\psi)(\tilde{\gamma}^\mu\partial_\mu\psi)-12i(m\psi)(\bar{\psi}\tilde{\gamma}^\mu\partial_\mu\psi)\\&+16(m\bar{\psi}\psi)(m\psi) ]+{\cal O}(\beta^2)=0.
\end{split}
\end{equation}
Since $\tilde{\gamma}^\mu\partial_\mu\psi=\gamma^\mu\partial_\mu+3i\beta\kappa^2(\bar{\psi}\gamma^\mu\partial_\mu\psi)\gamma^\mu\partial_\mu\psi-4\beta\kappa^2(m\bar{\psi}\psi)\gamma^\mu\partial_\mu\psi$, we transform Eq.~\eqref{118} back into the Jordan frame and, after some manipulations, we obtain the matter-field equation
\begin{equation}
\label{119}
\begin{split}
&(i\slashed{\partial}\psi-m\psi)[1+4\beta\kappa^2(3\bar{\psi}i\slashed{\partial}\psi-4m\bar{\psi}\psi)]+{\cal O}(\beta^2)\\&=(i\slashed{\partial}\psi-m\psi)[1+4\beta\kappa^2T]+{\cal O}(\beta^2)=0.
\end{split}
\end{equation}
Note that, once again, the matter equation in the Jordan frame can be written as $\phi^{-2}(i\slashed{\partial}\psi-m\psi)=0$. As in the previous case, the above equation has two solutions, namely $i\slashed{\partial}\psi-m\psi=0$ and $T={\rm const}$. The latter is not acceptable, as mentioned before, and thus the only viable solution is $i\slashed{\partial}\psi-m\psi=0$, which locally reduces to the Dirac equation at tree level. However, as in the case of the massless scalar field, due to the mixing between gravity and matter perturbations, at third order in perturbation theory the matter-field equation is affected by corrections that are lacking a $\kappa^2$ factor, thus is less suppressed with respect to other terms and can lead to measurable effects. Once again, we refer to the publicly available \textit{Mathematica} notebook at Ref.~\citep{Ventagli2023} for the explicit perturbed field equations.

\section{THE CASE OF $f({\cal R})={\cal R}-\mu^4/{\cal R}$}\label{App:SecondModel}

Although framed for simplicity for the case of $f({\cal R})={\cal R}+\beta {\cal R}^2$, the discussion of Sec.~\ref{Sec:MatterFields} applies in general. To confirm this statement, here we study the same Palatini $f(\cR)$ theory considered in Ref.~\citep{Flanagan:2003rb} and show that ---~in agreement with the latter reference~--- the field equations for matter fields in the Jordan frame are affected by corrections to the Standard Model that percolate from the gravity sector at third and higher orders in perturbation theory and are not Planckian suppressed. For simplicity, we focus only on the case of the massless scalar field, which suffices to make our point.

We focus on the expansion performed in Ref.~\citep{Flanagan:2003rb}, namely ${\mu^4}/{\mathcal{R}}\gg \mathcal{R}$ or, equivalently, $\mu^2/\kappa^2 T\gg 1$. This model admits de Sitter solutions also in the absence of a cosmological constant, with $\mu\propto H_0\simeq70.9({\rm km/s})/{\rm Mpc}$~\citep{Flanagan:2003rb}. Thus the expansion $\mu^2/\kappa^2 T\gg 1$ implies $T\ll 10^{-27}\text{kg/m}^3$, which is satisfied for late-time cosmological solutions.
It is clear that, under this expansion, we do not retrieve the Einstein equations as a limit of the theory.

To proceed with our analysis, we write the scalar field $\phi$ in terms of the trace of the stress-energy tensor. Since $\phi=f'(\mathcal{R})=1+\frac{\mu^4}{\mathcal{R}^2}$, we need to find the relation between the Ricci scalar and $T$. As in the previous case, from the algebraic Eq.~\eqref{algebraic}, we find $\mathcal{R}=\frac{1}{2}(-\kappa^2T+\sqrt{\kappa^4T^2+12\mu^4}\,)$, where we consider the positive sign of the root to allow for de Sitter solutions when $T=0$.
Substituting this solution into the definition of $\phi$ and expanding up to terms ${\cal O}(\mu^{-4})$ yields
\begin{equation}
\label{146b}
\phi=\frac{4}{3}+\frac{\kappa^2T}{3\sqrt{3}\mu^2}+\frac{\kappa^4T^2}{18\mu^4}+{\cal O}\left( \frac{1}{\mu^6} \right).
\end{equation}
We can now use this relation to integrate out the auxiliary scalar field from the field equations in the Jordan frame and from the action in the Einstein frame. Using Eq.~\eqref{146b}, we can rewrite the potential $V$ in terms of the matter source. Expanding and keeping terms up to ${\cal O}(\mu^{-4})$, we get
\begin{equation}
\label{146c}
V=\frac{2\mu^2}{\sqrt{3}}+\frac{\kappa^2T}{3}+\frac{\kappa^4T^2}{12\sqrt{3}\mu^2}+{\cal O}\left( \frac{1}{\mu^6} \right).
\end{equation}

With these relations at hand, we can derive the modified Einstein equation in the Jordan frame. Substituting Eqs.~\eqref{146b} and~\eqref{146c} into Eq.~\eqref{eq:Jordan} yields
\begin{equation}
\label{146e}
\begin{split}
G_{\mu\nu}&=\frac{3}{4}\kappa^2 T_{\mu\nu}-\frac{1}{16}g_{\mu\nu}\kappa^2 T-\frac{\sqrt{3}}{4}\mu^2g_{\mu\nu}\\&-\frac{\sqrt{3}}{16}\frac{\kappa^4TT_{\mu\nu}}{\mu^2}+\frac{1}{4\sqrt{3}}\frac{\kappa^2\nabla_\mu\nabla_\nu T}{\mu^2}\\&-\frac{1}{4\sqrt{3}}\frac{\kappa^2g_{\mu\nu}\square T}{\mu^2}+\frac{1}{64\sqrt{3}}\frac{\kappa^4g_{\mu\nu}T^2}{\mu^2}\\&+{\cal O}\left( \frac{1}{\mu^4}\right)\,.
\end{split}
\end{equation}

We now consider the modified Einstein equation in the Einstein frame. We first have to rewrite the potential $U$ in terms of the matter fields. Using $T=\phi^2\tilde{T}$ and the expression for $\phi$ derived in Eq.~\eqref{146b}, we obtain
\begin{equation}
\label{146f}
U=\frac{3\sqrt{3}\mu^2}{16\kappa^2}-\frac{\kappa^2\tilde{T}^2}{9\sqrt{3}\mu^2}+{\cal O}\left( \frac{1}{\mu^4} \right)\,.
\end{equation}

Using this relation we can write Eq.~\eqref{eqEF1} explicitly in terms of $\tilde{T}_{\mu\nu}$ and $\tilde{T}$, namely
\begin{equation}\label{146g}
\begin{split}
\tilde{G}_{\mu\nu}&=\kappa^2 \tilde{T}_{\mu\nu}-\frac{3\sqrt{3}}{16}\mu^2\tilde{g}_{\mu\nu}\\&+\frac{1}{9\sqrt{3}}\frac{\tilde{g}_{\mu\nu}\kappa^4 \tilde{T}}{\mu^2}+{\cal O}\left( \frac{1}{\mu^4} \right),
\end{split}
\end{equation}
which is dynamically equivalent to Eq.~\eqref{146e}, as expected.

Let us focus now on the matter sector, in order to check whether new matter interactions appear, as it was claimed in Ref.~\citep{Flanagan:2003rb}. For simplicity, we focus on the case of a massless scalar field. In the Einstein frame, the action reads
\begin{equation}
\label{146i}
\begin{split}
\tilde{S}&=\int d^4x \sqrt{-\tilde{g}} \big( \frac{\tilde{R}}{2\kappa^2}-\frac{3\sqrt{3}\mu^2}{16\kappa^2}+\frac{3}{8}\tilde{g}^{\mu\nu}\partial_\mu\psi\partial_\nu\psi \\&-\frac{\kappa^2\tilde{g}^{\mu\nu}\partial_\mu\psi\partial_\nu\psi\tilde{g}^{\lambda\sigma}\partial_\lambda\psi\partial_\sigma\psi}{16\sqrt{3}\mu^2}\big)+{\cal O}\left( \frac{1}{\mu^4} \right).
\end{split}
\end{equation} 
Varying the action with respect to the matter field $\psi$ yields
\begin{equation}
\label{146l}
\begin{split}
&\frac{3}{4}\tilde{\square}\psi-\frac{\kappa^2\tilde{g}^{\lambda\sigma}\partial_\lambda\psi\partial_\sigma\psi\tilde{\square}\psi}{4\sqrt{3}\mu^2}-\frac{\kappa^2\partial^\lambda\psi\partial^\sigma\psi\tilde{\nabla}_\lambda\partial_\sigma\psi}{2\sqrt{3}\mu^2}\\&+{\cal O}\left( \frac{1}{\mu^4} \right)=0.
\end{split}
\end{equation}
As before, we perform a conformal transformation to the Jordan frame. After some manipulations, the outcome is
\begin{equation}
\label{146m}
\begin{split}
&\square\psi\left( \frac{9}{16}-\frac{3\sqrt{3}}{32}\frac{\kappa^2\partial_\lambda\psi\partial^\lambda\psi}{\mu^2} \right)+{\cal O}\left( \frac{1}{\mu^4} \right)\\&=\square\psi\left( \frac{9}{16}-\frac{3\sqrt{3}}{32}\frac{\kappa^2T}{\mu^2} \right)+{\cal O}\left( \frac{1}{\mu^4} \right)\\
&=\phi^{-2}\square\psi=0.
\end{split}
\end{equation}
Again, the only acceptable solution to this equation is $\square\psi=0$, which is the massless scalar field equation in curved spacetime. However, performing an expansion around a flat background, one finds corrections to the leading order that appear from the third order onward and can be big, invalidating the perturbation scheme, unless one severely constrains the parameter of the theory. For example, at third order the leading term is corrected by a dimensionless term proportional to $\frac{\kappa^2}{\mu^2}\nabla\delta\psi^1\nabla\delta\psi^1$, which lacks a $\kappa^2$ suppression term. The perturbed field equations are publicly available at Ref.~\citep{Ventagli2023}.

One obtains a similar outcome even when considering a Dirac field, which was the case studied in Ref.~\citep{Flanagan:2003rb}. We do not report the analysis here, since the mechanism that leads to matter corrections is the same as the cases considered previously (the results are available at Ref.~\citep{Ventagli2023}). We only stress that when one maps these problematic terms to the Einstein frame, one retrieves the corrections found in Ref.~\citep{Flanagan:2003rb}.

%
\bibliographystyle{utphys}
\bibliography{Ref}

\end{document}